# Model underestimates of OH reactivity cause overestimate of hydrogen's climate impact

Laura H. Yang[1*], Daniel J. Jacob[1,2], Haipeng Lin[1], Ruijun Dang[1], Kelvin H. Bates[3], James D. East[1], Katherine R. Travis[4], Drew C. Pendergrass[1], Lee T. Murray[5]

[1] School of Engineering and Applied Sciences, Harvard University, Cambridge, MA, USA
[2] Department of Earth and Planetary Sciences, Harvard University, Cambridge, MA, USA
[3] Department of Mechanical Engineering, University of Colorado, Boulder, CO, USA
[4] NASA Langley Research Center, Hampton, VA, USA
[5] Department of Earth and Environmental Sciences, University of Rochester, Rochester, NY, USA
* Correspondence to: Laura Hyesung Yang (laurayang@g.harvard.edu)

**Deploying hydrogen technologies is one option to reduce energy carbon dioxide emissions, but recent studies have called attention to the indirect climate implications of fugitive hydrogen emissions. We find that biases in hydroxyl (OH) radical concentrations and reactivity in current atmospheric chemistry models may cause a 20% overestimate of the hydrogen Global Warming Potential (GWP). A better understanding of OH chemistry is critical for reliable estimates of the hydrogen GWP.**

There is growing interest in using hydrogen ($H_2$) technologies to decarbonize the economy[1]. However, fugitive emissions during the production, delivery, and use of hydrogen can contribute to the radiative forcing of climate[2]. Although hydrogen is not a greenhouse gas, its atmospheric oxidation by the hydroxyl radical (OH) increases the abundance of greenhouse gases, including methane (which competes with hydrogen for OH), tropospheric ozone, and stratospheric water vapor[3]. Atmospheric chemistry model studies have thus attributed a global warming potential (GWP) to hydrogen due to these indirect climate effects[4,5,6], but current models have known OH biases[5,6,7,8]. Here, we show that correcting these biases decreases the hydrogen GWP substantially.

Hydrogen is a well-mixed atmospheric gas with a lifetime of two years[9] and a present-day concentration of 530 ppbv. Fossil fuel combustion, biomass burning, and the oxidation of methane ($CH_4$) by OH are currently the dominant sources[9]. The sinks are uptake by soil and oxidation by OH[10].

The GWP of an emitted gas indicates the integrated radiative forcing over a certain time horizon from the emission of 1 kg of the gas relative to 1 kg of carbon dioxide ($CO_2$). Previous studies have used global 3-D atmospheric chemistry models to quantify the GWP of hydrogen due to contributions from methane, ozone, and stratospheric water vapor. Hauglustaine et al.[4] reported a hydrogen GWP over a 100-year (GWP-100) time horizon of 12.8±5.2 using GFDL-AM4.1. Warwick et al.[6] reported 12±6 using UKESM1. Sand et al.[5] reported 11.6±2.8. The magnitude of the soil sink is a major uncertainty in these estimates[10].

Observations of the methyl chloroform proxy imply a methane lifetime of 11.2±1.3 years against oxidation by tropospheric OH[11]. That lifetime is underestimated by 10-30% in current models because of excessive OH[5,6,7,8]. Accounting for UV absorption by water vapor could reduce model OH by 4%[12]. Models also underestimate the OH reactivity (OHR; loss frequency of OH), for which extensive measurements are available from surface sites and aircraft[13,14]. Reported model underestimates of OHR range from 30% in the remote troposphere[15] and 60% in polluted air[16] sampled by aircraft, to a factor of 2 to 10 in continental surface air[17,18]. Underestimate of OHR is commonly attributed to underestimate of carbon monoxide (CO)[19,20] and to nonmethane volatile organic compounds (NMVOCs) missing from the models[15,16,17,18], which implies that the models would overestimate the sensitivity of methane to hydrogen. A conceptual calculation presented in the SI shows that if a model underestimates OHR by a fraction $F$, the sensitivity of OH to hydrogen is overestimated by $(1-F)^{-1}-1$. For instance, underestimating OHR in the model by 30% leads to overestimating the sensitivity of OH to hydrogen by 43%.

Here, we use the GEOS-Chem global atmospheric chemistry model to investigate the effect of model OH and OHR biases in computing the GWP-100 of hydrogen. GEOS-Chem (https://geos-chem.org) is a widely used model that includes detailed tropospheric oxidant and halogen chemistry. Hydrogen in the GEOS-Chem model is prescribed as a constant mixing ratio of 500 ppbv. In this work, we treat hydrogen and methane emissions implicitly using monthly observed surface concentrations in different latitudinal bands as boundary conditions following Sand et al.[5]. Our comparison of the current standard version of GEOS-Chem to aircraft observations of OH and OHR (Figure S1) shows similar biases as previous GEOS-Chem studies[15,16]. To correct this OHR bias, we added NMVOC emissions to GEOS-Chem including per capita volatile chemical products (VCPs)[21,22] and oceanic alkanes[15]. This did not fully correct the OHR bias in the continental air, but we found that further addition of NMVOCs to close the bias resulted in excessive model ozone. We therefore corrected the remaining bias by adding a first-order OH sink in the continental troposphere, as described in the SI. This corrects the OH and OHR biases (Figure S1) without compromising other aspects of the simulation. The methane lifetime against oxidation by tropospheric OH rises to 11.4 years, compared to 8.5 years in the standard GEOS-Chem simulation, and is consistent with the observationally-based estimate of 11.2±1.3 years. We refer to that simulation as modified GEOS-Chem in what follows.

Table 1 shows the methane lifetime in GEOS-Chem and in previous hydrogen GWP model studies, together with estimates from IPCC AR6[23] which adopted the observationally-based estimate in its calculation of the methane GWP[11]. There is no hydrogen GWP assessment in IPCC AR6. All previous model studies examining the hydrogen GWP underestimate methane lifetime by 15-26%, and so does the standard GEOS-Chem. Also shown in the Table is the methane feedback factor $f$, defined as the ratio of the perturbation lifetime to the total atmospheric lifetime, which diagnoses the positive feedback from decreased OH due to methane addition in the GWP calculation. Previous model studies and the standard GEOS-Chem give $f$ values in the range of 1.39-1.45, higher than the value 1.30±0.07 recommended by IPCC AR6[23] and modified GEOS-Chem ($f$ =

1.31), implying excessive $CH_4$-OH coupling as would result from an OHR underestimate. Previous models and the standard GEOS-Chem underestimate the methane lifetime and overestimate the feedback factor, which partially offset each other in the effect on the perturbation lifetime. The perturbation lifetime is used for methane GWP estimates and it ranges from 10.4 to 11.8 years in the previous model studies and the standard GEOS-Chem, as compared to the IPCC AR6 recommendation of 12.5±1.8 years and 12.8 years in the modified GEOS-Chem.

The global tropospheric ozone burden in the modified GEOS-Chem (359 Tg) and the standard GEOS-Chem (346 Tg) are within the 347±28 Tg range reported by IPCC AR6[23]. Ozone in the modified GEOS-Chem is slightly higher because of the added NMVOCs, more than compensating for the added first-order OH sink (Table S1).

We follow the Sand et al.[5] method to calculate the hydrogen GWP-100 as the sum of the contributions from methane, ozone, and stratospheric water vapor. Perturbations to hydrogen and methane emissions are applied separately by increasing the surface concentrations used as boundary conditions by 10% and inferring the corresponding changes in emissions by atmospheric mass balance. We use a hydrogen soil sink of 59 Tg $a^{-1}$ to compute hydrogen lifetime and compare to the Sand et al.[5] results using the same soil sink. We spin up the GEOS-Chem simulations for three years with 2016 meteorological data to achieve a steady state in the responses of atmospheric hydrogen, ozone, and methane concentrations to the perturbations, and we report results for the fourth year. Hydrogen and methane respond rapidly to the perturbations in these simulations because their surface concentrations are imposed. We use the same hydrogen GWP-100 from stratospheric water vapor as Sand et al.[5] because its computation would require a much longer model spin-up and is not sensitive to OHR and OH bias corrections.

Figure 1 compares the hydrogen GWP-100 values from Sand et al.[5] and those computed using the standard and modified GEOS-Chem. Sand et al.[5] and the standard GEOS-Chem have similar total hydrogen GWP-100 values of 11.4±1.9 and 10.8, respectively. The modified GEOS-Chem shows a hydrogen GWP-100 of 8.8, 20% lower than the standard GEOS-Chem, due to reduced methane (17%) and ozone (31%) contributions. The reduced methane contribution is because OH is less sensitive to hydrogen perturbation in the modified model. The reduced ozone contribution reflects the greater importance of NMVOCs in driving the ozone production in the modified model; because of this, the response of ozone to a change in hydrogen or methane is weaker. The effect is compounded by the weaker response of methane to the hydrogen perturbation.

In summary, we have shown that current biases in global atmospheric chemistry model simulations of OH concentrations and OH reactivity (OHR) lead to overestimates of the calculated global warming potential (GWP) for hydrogen. The hydrogen GWP-100 decreases by 20% in the GEOS-Chem model when these biases are corrected by adding non-methane volatile organic compounds (NMVOCs) emissions and a first-order sink for OH. Our resulting best estimate for the hydrogen GWP-100 is 8.8. Better understanding of NMVOCs emissions, tropospheric OH chemistry, and the hydrogen soil sink are needed to obtain a more precise hydrogen GWP.

**Acknowledgment**. The authors acknowledge funding support from the ExxonMobil Technology and Engineering Company and a National Science Foundation Graduate Research Fellowship under grant no. DGE 2140743. We thank Bryan K. Mignone, Emily K. Reidy, Olivia E. Clifton, Ragnhild Bieltvedt Skeie, Loretta J. Mickley, and Todd A. Mooring for the helpful discussion. The views expressed in this paper are those of the authors alone.

**Data availability.** The GEOS-Chem simulation results are available upon request from the corresponding author.

**Code availability.** The GEOS-Chem model code is open source (https://doi.org/10.5281/zenodo.10640383). The calculation used for hydrogen GWP-100 follows Sand et al.[5] work and is publicly available in the form of jupyter notebooks as version v0.1.0 of this code https://github.com/ciceroOslo/Hydrogen_GWP on github, released under the Apache-2.0 license.

**Author contributions.** LHY and DJJ contributed to the study conceptualization. LHY conducted the data and modeling analysis with contributions from DJJ, HL, RD, KHB, JDE, KRT, DCP, and LTM. LHY and DJJ wrote the paper with input from all the co-authors.

**Competing interests.** The authors declare no competing interests.

**Supplementary information.** SI provides a calculation of how underestimating OH reactivity leads to overestimating the sensitivity of methane to hydrogen, model description, tropospheric ozone budget analysis, calculations of hydrogen global warming potential from ozone and methane, and vertical profiles of OH and OH reactivity.

**References.**

**Table 1. Methane lifetimes in models applied to hydrogen GWP calculations**.

|  | Lifetime against tropospheric OH (years) | Total atmospheric lifetime (years)[a] | Perturbation lifetime (years)[b] | Methane feedback factor[c] |
|---|---|---|---|---|
| Sand et al.[5] | 8.3±0.9[d] | 7.2±0.9 | 10.4±0.9 | 1.45±0.07 |
| Hauglustaine et al.[4] | 9.5[d] | 8.4 | 11.8 | 1.40 |
| Warwick et al.[6] | 8.5 | 7.6 | 10.4 | 1.39 |
| Standard GEOS-Chem | 8.5 | 7.5 | 10.7 | 1.42 |
| Modified GEOS-Chem | 11.4 | 9.8 | 12.8 | 1.31 |
| IPCC AR6[23] | 11.2±1.3[e] | 9.6±0.9[f] | 12.5±1.8[f] | 1.30±0.07[g] |

[a]: Accounting for additional minor sinks from stratospheric oxidation (120 years) and soil uptake (160 years) as used by Sand et al.[5].
[b]: e-folding time scale for the decay of an instantaneous addition of methane as used in methane GWP calculations.
[c]: Ratio of the perturbation lifetime to the total atmospheric lifetime, reflecting the decrease in OH from a methane addition.
[d]: These studies only give the methane lifetime against oxidation by total (tropospheric + stratospheric) OH. In GEOS-Chem, the methane lifetime against tropospheric OH is 0.25 years longer than against total OH and we apply that correction to the numbers given in the original publications.
[e]: Based on methyl chloroform observations[11].
[f]: IPCC AR6 reports 9.1±0.9 years for the total atmospheric lifetime and 11.8±1.8 years for the perturbation lifetime, but with different values for the additional minor sinks in footnote a. We adjust these values here to match the magnitudes of the minor sinks in footnote a.
[g]: Based on a six-member ensemble of AerChemMIP ESMs with a pre-industrial baseline.

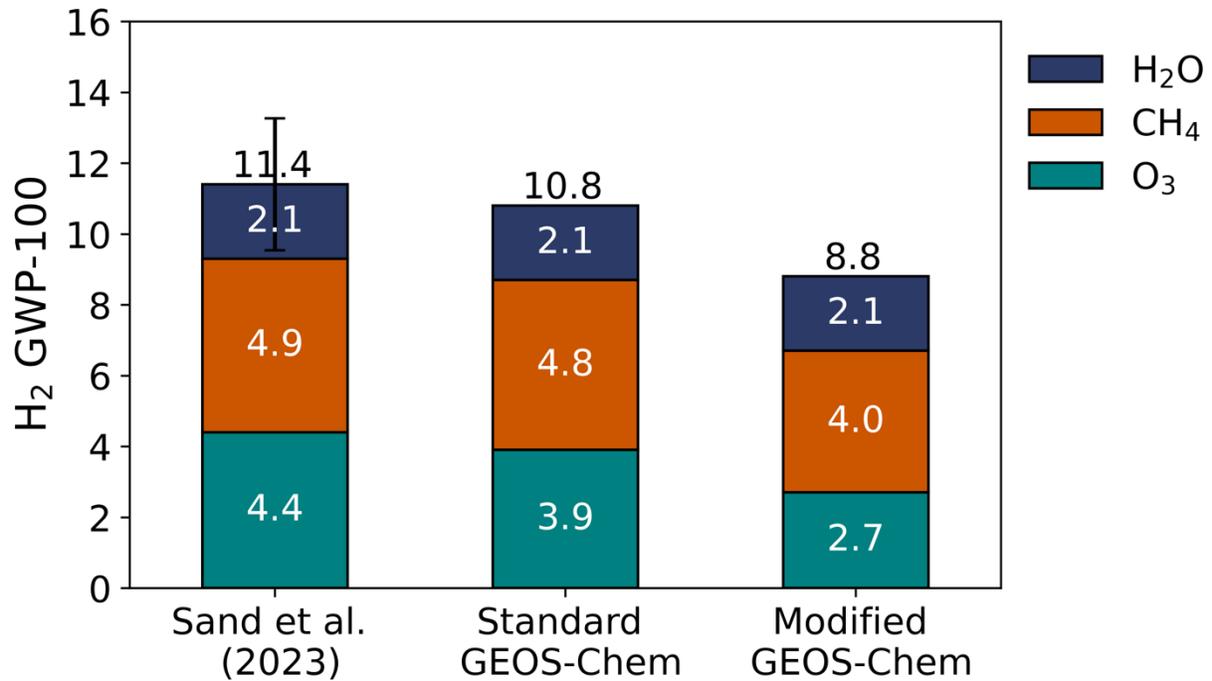

**Figure 1. Global Warming Potential of hydrogen over 100-year horizon (GWP-100)**. The Figure compares results from the standard GEOS-Chem model, the modified GEOS-Chem model, and the mean of five models reported by Sand et al.[5]. Hydrogen GWP-100 is expressed as the sum of contributions from ozone, methane, and stratospheric water vapor. The GEOS-Chem simulations use the same stratospheric water vapor contribution and the same hydrogen soil sink of 59 Tg a$^{-1}$ as in Sand et al.[5]. The vertical error bar on the Sand et al.[5] estimate is obtained by applying their error propagation method but excluding the uncertainty from the soil sink, as all models use the same soil sink value. Sand et al.[5] present a larger one-standard deviation uncertainty due to soil sink.

## Supplementary Information for

**Model underestimates of OH reactivity cause overestimate of hydrogen's climate impact**

Laura H. Yang[1*], Daniel J. Jacob[1,2], Haipeng Lin[1], Ruijun Dang[1], Kelvin H. Bates[3], James D. East[1], Katherine R. Travis[4], Drew C. Pendergrass[1], Lee T. Murray[5]

[1] Harvard John A. Paulson School of Engineering and Applied Sciences, Harvard University, Cambridge, MA, USA
[2] Department of Earth and Planetary Sciences, Harvard University, Cambridge, MA, USA
[3] Department of Mechanical Engineering, University of Colorado, Boulder, CO, USA
[4] NASA Langley Research Center, Hampton, VA, USA
[5] Department of Earth and Environmental Sciences, University of Rochester, Rochester, NY, USA
* Correspondence to: Laura Hyesung Yang (laurayang@g.harvard.edu)

**This PDF file includes:**

- Methods
  - Calculation of how underestimating OH reactivity leads to overestimating the sensitivity of methane to hydrogen
  - Model description
- Tropospheric ozone budget analysis
  - Table S1. Global tropospheric ozone budget using odd oxygen family ($O_x$)
- Calculation of hydrogen global warming potential from methane
- Calculation of hydrogen global warming potential from ozone
  - Table S2. Non-methane-induced and methane-induced hydrogen absolute global warming potential from ozone in standard and modified GEOS-Chem
- Figure S1: Vertical profiles of hydroxyl radical (OH) and OH reactivity
- SI References

# Methods

**Calculation of how underestimating OH reactivity leads to overestimating the sensitivity of methane to hydrogen.**

At a steady state, the production rate of OH ($P_{OH}$) equals the loss rate of OH ($L_{OH}$).

$$P_{OH} = L_{OH} \qquad \text{eq. 1}$$

$L_{OH}$ can be expressed as OH reactivity (k) times the concentration of OH ([OH]).

$$L_{OH} = k[OH] \qquad \text{eq. 2}$$

where $k = k_{H_2+OH}[H_2] + k_{CH_4+OH}[CH_4] + k_{CO+OH}[CO] + \sum k_{VOC+OH}[VOC] + \cdots$

If we substitute equation 2 into equation 1, we can express the [OH] as the following:

$$[OH] = \frac{P_{OH}}{k} \qquad \text{eq. 3}$$

Take the derivative of OH with respect to $H_2$.

$$\frac{d[OH]}{d[H_2]} = \frac{d}{d[H_2]}\left(\frac{P_{OH}}{k}\right) \qquad \text{eq. 4}$$

Apply the quotient rule to equation 4 and obtain equation 5. This equation shows the sensitivity of OH to $H_2$.

$$\frac{d[OH]}{d[H_2]} = -\frac{k_{H_2+OH} P_{OH}}{k^2} \qquad \text{eq. 5}$$

Assume that k is underestimated in the model by a fraction *F* due to missing species compared to the real atmosphere. This means that $P_{OH}$ from the model is underestimated by a fraction *F* from equation 1. Plugging into the sensitivity equation 5, the model will overestimate the sensitivity of OH to hydrogen by a fraction $(1-F)^{-1}-1$.

**Model description**

The NASA aircraft campaigns provided an opportunity to evaluate GEOS-Chem against measured atmospheric composition, including OH and OH reactivity (OHR; Brune et al., 2020). The ATom-1 campaign took place from July to August 2016 over the Pacific and Atlantic Oceans, measuring the remote air (Brune et al., 2020). The KORUS-AQ campaign occurred over the Korean peninsula from May to June 2016, measuring polluted air (Crawford et al., 2021). We used ATom-1 and KORUS-AQ campaigns to evaluate our simulation results, as they provided OH and OHR measurements and took place in the same year as our model runs. The Airborne Tropospheric Hydrogen Oxides Sensor (ATHOS) measured OH and OHR in ATom-1 and KORUS-AQ campaigns (Faloona et al., 2004; Mao et al., 2009).

We used GEOS-Chem chemical transport model (CTM) version 14.3.0 (v14.3.0; https://doi.org/10.5281/zenodo.10640536) driven by assimilated meteorological data from MERRA-2. For all analyses related to methane lifetime, feedback factor, and hydrogen global warming potential (GWP), we used a global 4°×5° simulation with 47 vertical layers. We conducted 0.5°×0.625° nested simulation, using boundary conditions from the 4°×5° simulation, for comparison with the KORUS-AQ aircraft campaign. Outputs from the 4°×5° simulation were compared with the ATom-1 aircraft campaign (Figure S1).

We established two model configurations to investigate the impact of underestimated OHR on the computation of hydrogen GWP: the standard GEOS-Chem version 14.3.0 and the modified GEOS-Chem with increased OHR. The modified GEOS-Chem better matched the observational constraints from aircraft measurements (Figure S1) and the methane lifetime against tropospheric OH inferred from methyl chloroform observations (Table 1). The OHR in the modified GEOS-Chem was increased by 1) adding the additional emissions from non-methane volatile organic compounds (NMVOCs) and 2) prescribing the missing OHR over continents.

We added global non-industrial volatile chemical products (VCPs) emission scaled by population density, following Yang et al. (2023). Additionally, we incorporated oceanic emissions of alkanes (Travis et al., 2020) to match the OHR observed during the ATom-1 aircraft campaign. These emission adjustments helped with near-boundary layer OHR but could not reconcile the missing OHR in the troposphere compared to the KORUS-AQ aircraft observation. Also, further adding NMVOCs in GEOS-Chem at the levels needed to correct OHR led to excessive ozone through OH recycling. Hence, we prescribed a missing OHR value of 1 $s^{-1}$ in the boundary layer and 1/6 $s^{-1}$ in the free troposphere in the continental air to fully reconcile the missing OHR compared to the KORUS-AQ aircraft observation.

## Tropospheric ozone budget analysis

We use the odd oxygen family ($O_x$) to understand the tropospheric ozone budget (Table S1). The $O_x$ family is defined as follows:

$$O_x \equiv O_3 + O + O(^1D) + NO_2 + 2NO_3 + 3N_2O_5 + HNO_3 + HNO_4 + PANs.$$

**Table S1. Global tropospheric ozone budget using odd oxygen family ($O_x$).**

|  | Standard GEOS-Chem (CTRL) | Modified GEOS-Chem (CTRL) |
|---|---|---|
| Tropospheric burden | | |
| $O_3$ burden (Tg $O_3$) | 346 | 359 |
| $HO_2$ burden (Gg $HO_2$) | 26.2 | 25.8 |
| $O_x$ chemical sources (Tg $O_3$ a$^{-1}$) | | |
| $NO + HO_2 \rightarrow NO_2 + OH$ | 3970 | 3593 |
| $CH_3O_2 + HO_2 \rightarrow CH_3OOH + O_2$ | 1418 | 1316 |
| $NO + RO_2 \rightarrow NO_2 + RO$ | 330 | 903 |
| Total $O_x$ chemical source | 5718 | 5812 |
| $O_x$ chemical sinks (Tg $O_3$ a$^{-1}$) | | |
| $O_3 + HO_2 \rightarrow OH + O_2$ | 1337 | 1357 |
| $O_3 + OH \rightarrow HO_2 + O_2$ | 663 | 498 |
| $O^1D + H_2O \rightarrow 2OH$ | 2169 | 2290 |
| Other losses | 1047 | 1138 |
| Total $O_x$ chemical sink | 5216 | 5283 |
| Lifetime (days) | | |
| Lifetime of $O_x$ | 24.2 | 24.8 |

The added non-methane volatile organic compounds (NMVOCs) increase the organic peroxy radical ($RO_2$) burden in the modified GEOS-Chem, increasing the production of ozone ($O_3$) via the $NO + RO_2$ reaction pathway. The added first-order OH sink decreases the hydroperoxyl radical ($HO_2$) burden in the modified GEOS-Chem, reducing $O_3$ production through the $NO + HO_2$ and $CH_3O_2 + HO_2$ reaction pathways. Since the increase in $O_3$ production via the $NO + RO_2$ reaction pathway is greater than the decrease in $O_3$ production via the $NO + HO_2$ and $CH_3O_2 + HO_2$ reaction pathways, modified GEOS-Chem has a higher $O_3$ burden than the standard GEOS-Chem.

## Calculation of hydrogen global warming potential from methane

Hydrogen ($H_2$) absolute global warming potential (AGWP) from methane ($CH_4$) per 1 Tg $a^{-1}$ emission of $H_2$ (mW $m^{-2}$ a $Tg(H_2)^{-1}$) can be calculated following equation 6.

$$AGWP_{H_2 \text{ from } CH_4} = r_{CH_4} \cdot \frac{CH_4 \text{ surface conc.}(\Delta CH_4)}{CH_4 \text{ flux }(\Delta CH_4)} \cdot \frac{CH_4 \text{ flux }(\Delta H_2)}{H_2 \text{ flux }(\Delta H_2)} \qquad \text{eq. 6}$$

where $r_{CH_4}$ is $CH_4$ radiative efficiency (0.448 mW $m^{-2}$ $ppb^{-1}$; Etminan et al., 2016) adjusted by -14% (Forster et al., 2021). The $CH_4$ flux is calculated by dividing the $CH_4$ burden by the total lifetime of $CH_4$ in each model run. The $H_2$ flux is calculated by dividing the $H_2$ burden by the total lifetime of $H_2$. $\Delta CH_4$ denotes the change between the $CH_4$ perturbation run and the control run. $\Delta H_2$ denotes the change between the $H_2$ perturbation run and the control run. The multiplication of the second and the third terms in equation 6 represents an increase in $CH_4$ surface concentration per 1 Tg $a^{-1}$ emission of $H_2$ that would have occurred if $CH_4$ surface concentration had not been kept fixed in the $H_2$ perturbation run.

The change in $CH_4$ flux ($Tg(CH_4)$ $a^{-1}$) between the $CH_4$ perturbation run and the control run can be expressed as equation 7.

$$CH_4 \text{ flux }(\Delta CH_4) = \frac{Burden_{CH_4}(\Delta CH_4)}{ff_{CH_4} \cdot \tau_{CH_4}(CTRL)} \qquad \text{eq. 7}$$

where $ff_i$ represents the feedback factor of species i, and $\tau_i(CTRL)$ denotes the lifetime of species i in the control run.

The change in $H_2$ flux ($Tg(H_2)$ $a^{-1}$) between the $H_2$ perturbation run and the control run can be expressed as equation 8.

$$H_2 \text{ flux }(\Delta H_2) = \frac{Burden_{H_2}(\Delta H_2)}{ff_{H_2} \cdot \tau_{H_2}(CTRL)} \qquad \text{eq. 8}$$

Once we plug equations 7 and 8 into equation 6, we can express the $H_2$ AGWP from $CH_4$ per 1 Tg $a^{-1}$ emission of $H_2$ (mW $m^{-2}$ a $Tg(H_2)^{-1}$) as:

$$AGWP_{H_2 \text{ from } CH_4} = r_{CH_4} \cdot \frac{ff_{CH_4} \cdot \tau_{CH_4}(CTRL) \cdot CH_4 \text{ surface conc.}(\Delta CH_4)}{Burden_{CH_4}(\Delta CH_4)} \cdot \frac{ff_{H_2} \cdot \tau_{H_2}(CTRL) \cdot CH_4 \text{ flux }(\Delta H_2)}{Burden_{H_2}(\Delta H_2)} \qquad \text{eq. 9}$$

We divide the $H_2$ AGWP from $CH_4$ by AGWP-100 $CO_2$ (0.0895 mW $m^{-2}$ a $Tg^{-1}$) from the IPCC AR6 report to obtain the $H_2$ GWP-100 from $CH_4$. Standard GEOS-Chem and modified GEOS-Chem have different feedback factors, lifetimes in the control run, and $CH_4$ flux($\Delta H_2$).

## Calculation of hydrogen global warming potential from ozone

Hydrogen ($H_2$) absolute global warming potential (AGWP) from ozone ($O_3$) is composed of non-$CH_4$-induced and $CH_4$-induced components. Equation 10 defines the non-$CH_4$-induced $H_2$ AGWP from $O_3$ per 1 Tg a$^{-1}$ emission of $H_2$ (mW m$^{-2}$ a Tg($H_2$)$^{-1}$).

$$\text{AGWP}_{H_2 \text{ from } O_3 \text{ not induced by } CH_4} = \frac{\text{ERF}_{O_3}(\Delta H_2)}{H_2 \text{ flux } (\Delta H_2)} \quad \text{eq. 10}$$

where $H_2$ flux ($\Delta H_2$) is the change in $H_2$ flux (Tg($H_2$) a$^{-1}$) between the $H_2$ perturbation run and the control run (eq. 8). $\text{ERF}_{O_3}(\Delta H_2)$ is $O_3$ effective radiative forcing (ERF) caused by $H_2$ perturbation (mW m$^{-2}$). $\text{ERF}_{O_3}(\Delta H_2)$ is calculated by multiplying the change in $O_3$ concentration between the $H_2$ perturbation run and the control run by the radiative kernel from Skeie et al. (2020).

Equation 11 defines the $CH_4$-induced $H_2$ AGWP from $O_3$ per 1 Tg a$^{-1}$ emission of $H_2$ (mW m$^{-2}$ a Tg($H_2$)$^{-1}$).

$$\text{AGWP}_{H_2 \text{ from } O_3 \text{ induced by } CH_4} = \frac{\text{ERF}_{O_3}(\Delta CH_4)}{CH_4 \text{ surface conc.}(\Delta CH_4)} \cdot \frac{CH_4 \text{ surface conc. } (\Delta CH_4)}{CH_4 \text{ flux } (\Delta CH_4)} \cdot \frac{CH_4 \text{ flux } (\Delta H_2)}{H_2 \text{ flux } (\Delta H_2)} \quad \text{eq. 11}$$

where $\text{ERF}_{O_3}(\Delta CH_4)$ is ozone ERF induced by $CH_4$ perturbation (mW m$^{-2}$). We calculate $\text{ERF}_{O_3}(\Delta CH_4)$ by multiplying the change in $O_3$ concentration between the $CH_4$ perturbation run and the control run by the radiative kernel from Skeie et al. (2020). The multiplication of the second and the third terms in equation 11 represents an increase in $CH_4$ surface concentration per 1 Tg a$^{-1}$ emission of $H_2$ that would have occurred if $CH_4$ surface concentration had not been kept fixed in the $H_2$ perturbation run. They are multiplied to normalize the result per 1 Tg a$^{-1}$ emission of $H_2$.

The total $H_2$ AGWP from $O_3$ is calculated by summing the non-$CH_4$-induced component (eq. 10) and $CH_4$-induced component (eq. 11) of the $H_2$ AGWP from $O_3$ (Table S2). Finally, we divide the total $H_2$ AGWP from $O_3$ by AGWP-100 $CO_2$ (0.0895 mW m$^{-2}$ a Tg$^{-1}$) from the IPCC AR6 report to obtain the $H_2$ GWP-100 from $O_3$.

**Table S2. Non-methane-induced and methane-induced hydrogen absolute global warming potential from ozone in standard and modified GEOS-Chem.**

|  | Standard GEOS-Chem | Modified GEOS-Chem |
|---|---|---|
| Unit: mW m$^{-2}$ a Tg($H_2$)$^{-1}$ | | |
| Non-$CH_4$-induced $H_2$ AGWP from $O_3$ | 0.165 | 0.137 |
| $CH_4$-induced $H_2$ AGWP from $O_3$ | 0.188 | 0.107 |
| Total $H_2$ AGWP from $O_3$ | 0.353 | 0.244 |

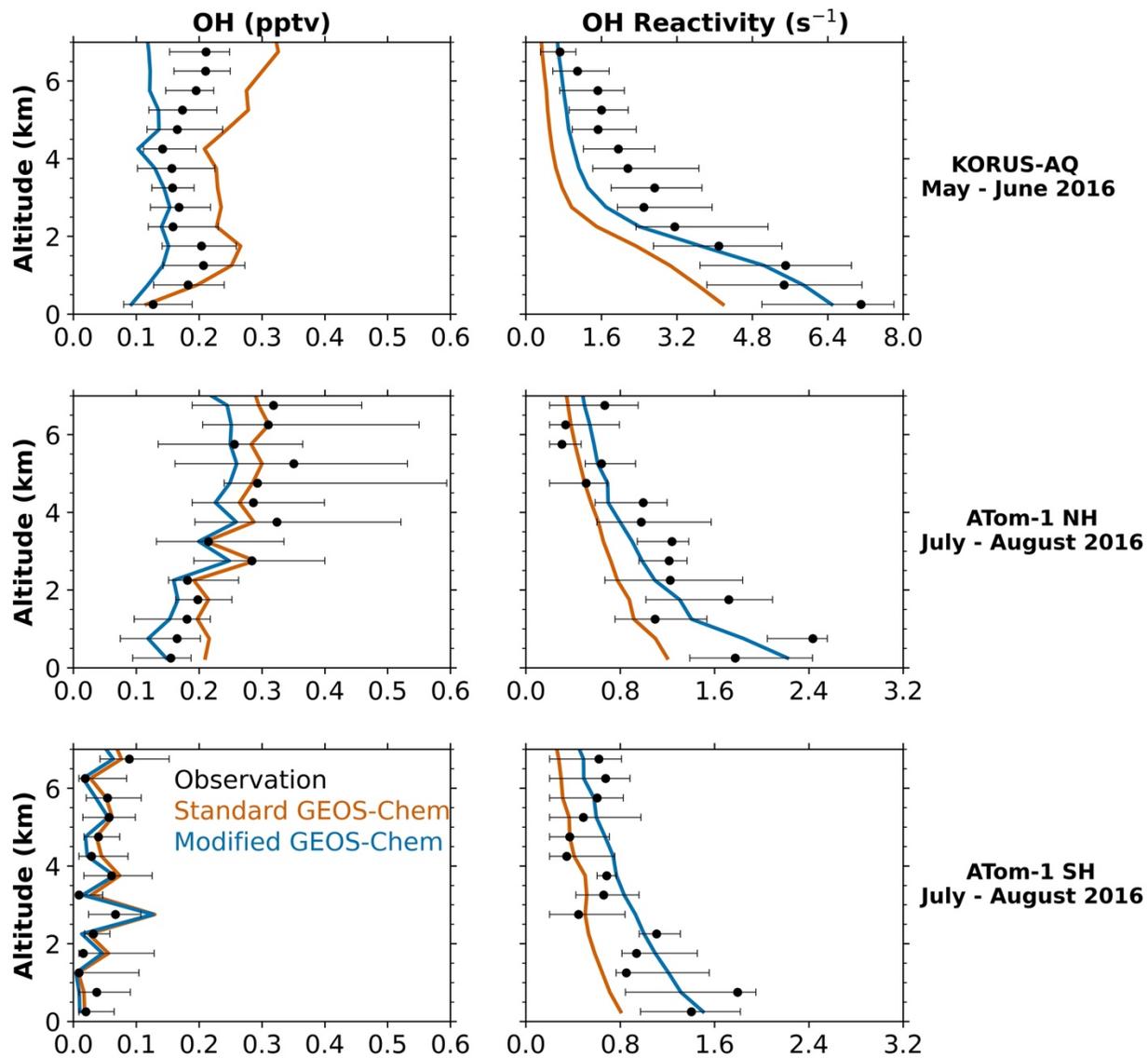

**Figure S1. Vertical profiles of hydroxyl radical (OH) and OH reactivity.** Median vertical profiles of OH and OHR from the KORUS-AQ aircraft campaign (Crawford et al., 2021) in May – June 2016 over the Seoul Metropolitan Area (SMA; 37 – 37.6°N, 126.6 – 127.7°E) and ATom-1 aircraft campaign (Brune et al., 2020) in July – August 2016 over Pacific and Atlantic oceans. Aircraft observations are compared to standard GEOS-Chem and modified GEOS-Chem with increased OHR. The error bar shows the 25$^{th}$ – 75$^{th}$ percentile range of the observation.